\documentclass[12pt]{article}

\def\be{\begin{equation}}
\def\ee{\end{equation}}
\def\beq{\begin{eqnarray}}
\def\eeq{\end{eqnarray}}

\def\p{\partial}

\def\s{\sigma}
\def\t{\tilde}
\def\vp{\varphi}

\def\N{{\cal N}}

\def\({\left (}
\def\){\right )}
\def\[{\left [}
\def\[{\right ]}

\def\ra{\rightarrow}

\begin{document}

\begin{titlepage}
\bigskip
\rightline{}
\rightline{gr-qc/0410049}
\bigskip\bigskip\bigskip\bigskip
\centerline
{\Large \bf {Spacetime in String Theory\footnote{To appear in ``Spacetime 100
Years Later", eds J. Pullin and R. Price (2005).}}}
\bigskip\bigskip
\bigskip\bigskip

\centerline{\large Gary T. Horowitz}
\bigskip\bigskip
\centerline{\em Department of Physics, UCSB, Santa Barbara, CA 93106}
\centerline{\em gary@physics.ucsb.edu}
\bigskip\bigskip

\begin{abstract}
We give a brief overview of the nature of spacetime emerging from
string theory. This is radically different from the
familiar spacetime of Einstein's relativity. At a perturbative level,
the spacetime metric appears as ``coupling constants" in a two
dimensional quantum field theory. Nonperturbatively (with
certain boundary conditions), spacetime is not fundamental but must be
reconstructed from a holographic, dual theory.

\end{abstract}

\end{titlepage}

\baselineskip=18pt

\setcounter{equation}{0}
\section{Introduction}
A hundred years ago our view of space and time was dramatically changed
by the introduction of special relativity. Ten years after that, Einstein made
spacetime dynamical in his general theory of relativity.
It has long been expected
that quantum gravity will require an even more radical change in our view
of spacetime. 
String theory is a promising approach to a consistent quantum theory of gravity.
In the past few decades a new picture of spacetime has been emerging from
this theory. While this picture is far from complete, it is already
clear that spacetime has many different features than it does in  
 relativity. I will discuss some of these new features below. (For a 
 description of spacetime in another approach to quantum gravity, see
 \cite{Ashtekar}.) This will
 be a nontechnical discussion focusing on the main ideas and results.
 More details can be found in the references which hopefully provide
 an introduction to the (vast) literature (see also \cite{Marolf:2003tx}).
 
String theory starts with the idea that fundamental particles are not
point-like but excitations of a one dimensional string. These strings have
a tension which defines a new fundamental length scale in the theory $\ell_s$.
The first thing one notices when one quantizes a string in flat spacetime is
that one needs more than four spacetime dimensions. The second thing is
that the ground state of the string is a tachyon. To remove the tachyon,
one adds fermions and requires that the string be supersymmetric. This
superstring is consistent in ten spacetime dimensions.

Thus string theory incorporates two major changes to the spacetime of
general relativity that were
proposed long  before string theory became popular in the mid 1980's.
The fact that spacetime might have more than four dimensions goes back to
Kaluza and Klein in the 1920's. The standard explanation for why
we have not seen these extra dimensions is that they are wrapped up in a 
small compact manifold. Supersymmetry is usually described as a 
symmetry between bosons and fermions, but it is much more than just another 
symmetry of the matter. It is really an extension of the Poincare symmetry
of spacetime, and can be viewed as saying that there are fermionic
directions as well as bosonic directions to spacetime.  In this
sense, it
describes the first extension of spacetime since space and time were unified by
Minkowski\footnote{Actually, supersymmetry was first developed in the context
of two dimensional 
string worldsheets, and later extended to four dimensional field theories
and gravity.}.

However this is just the beginning of the story.
Spacetime in string theory is not just
the usual spacetime with a few extra (bosonic and fermionic) dimensions.
As we will see, there are symmetries which equate geometrically different
spacetimes and even topologically different spacetimes. There are ways to 
resolve spacetime singularities. There are also ways to
reduce the dimension of spacetime, and in some cases, eliminate it altogether.

To begin our discussion, we remind the reader of
a few basic facts about string theory\footnote{By now there are several 
excellent textbooks on string theory, e.g., \cite{POL,ZEI}.}.
If one quantizes a free relativistic (super)string in flat
spacetime one finds a infinite tower of modes of increasing mass. 
Let us assume the string is closed, i.e., topologically a circle.
Then at the massless level, there is a symmetric traceless tensor mode
 which is identified with a linearized graviton. There is also an
 antisymmetric  tensor mode $B_{\mu\nu}$, and a scalar $\phi$ called the dilaton.
These states arise even without supersymmetry. Bilinears in the fermions
produce additional massless bosonic states which are higher rank
generalizations of Maxwell fields. They  are described by $p$-forms
$F_p=dA_{p-1}$.

Next, one postulates a simple splitting and joining interaction between strings.
The strength of this interaction is given by a dimensionless coupling constant
$g_s$ (which is related to  the dilaton).
Newton's constant $G$ is not an independent parameter,
but given in terms of $g_s$ and
$\ell_s$ by $G\sim g_s^2 \ell_s^8$ in ten dimensions.
Remarkably, one can show that this simple splitting and joining interaction
 reproduces the perturbative expansion of general
relativity. This was the earliest indication that string theory
incorporated general relativity. But string theory is certainly not
restricted to perturbing about flat spacetime. 
We will see how to recover the full vacuum Einstein equation,
$R_{\mu\nu}=0$, directly from string theory in the next section.

Over the past decade, it has become clear that that string theory is much
more than just a theory of strings. There are other extended objects called
branes. The name comes from membranes which are two dimensional, but branes
exist in any dimension: 0-branes are point particles, 1-branes are strings,
etc. Branes are nonperturbative objects in that their tension is
inversely related to a power of the coupling $g_s$. The most common type
of brane is called a D-brane and it has a tension $T\propto 1/g_s$. So
one would never
see these objects in perturbation theory in $g_s$. Even though they are very
heavy,  the gravitational field they produce is governed by $GT\sim g_s$
so as $g_s\rightarrow 0$, there should be a flat space description of these
objects and it was found by Polchinski \cite{Polchinski:1995mt}.
At weak coupling, a D-brane is a 
surface in Minkowski spacetime on which open strings can end. The D 
stands for ``Dirichlet" and refers to the boundary conditions on the  ends
of the open 
strings. These open strings move freely along the brane but cannot leave the
brane unless the endpoints join and form a closed string. The
massless states of an open string include a spin one particle, so
every D-brane comes with a $U(1)$ gauge field. When $N$ D-branes
come together, the open strings stretching from one to another also
become massless. This enhances the resulting gauge group from $U(1)^N$
to $U(N)$. These
D-branes are also sources for the $p$-form fields $F_p$.

The idea that there can be extended objects with degrees of freedom stuck
to the brane has given rise to an entirely new way to view extra
spatial dimensions, called brane-worlds 
\cite{Arkani-Hamed:1998rs,Randall:1999vf}. Rather than imagine that the
extra dimensions are very small, one can imagine that they are much larger
and we live on a $3+1$ dimensional brane in this higher dimensional
space. In other words, all observed
particles (quarks, leptons,  gauge bosons, etc.) are confined to move
on the brane and only gravity exists in the bulk since the graviton
comes from closed strings. One important consequence is that the
fundamental higher dimensional Planck scale could be of order a TeV,
 and we might see quantum gravity or string theory effects at the Large
 Hadron Collider at CERN.
This would be tremendously exciting, but it is important to remember that
this is only one of many possible scenarios and it not a unique prediction
of string theory.

The fact that the extra dimensions can be relatively large, has motivated
a study of black holes in more than four spacetime dimensions. This has
resulted in 
a number of surprises including a demonstration that stationary
solutions with event horizons are not characterized by their mass, charge and
angular momentum \cite{Emparan:2001wn}:
There is a violation of black hole uniqueness in higher dimensions. Among
the new stationary solutions that have recently been found,
there are neutral black rings \cite{Emparan:2001wn}, 
supersymmetric charged black rings \cite{Elvang:2004rt},
and nonuniform black strings \cite{Wiseman:2002zc}.
I will not discuss these further since they are basically a result of
general relativity in higher dimensions. I want to focus on the more 
fundamental changes in spacetime that arise from string theory.

In the next section, I discuss the role of spacetime in perturbative 
string theory, and indicate how different geometries and topologies can be
equivalent. Section three contains a discussion of perhaps the most important
lesson about spacetime that has emerged from nonperturbative string theory:
holography. The last section contains some concluding remarks.

\setcounter{equation}{0}
\section{Perturbative probes of spacetime}

At the perturbative level, one starts with a background spacetime metric 
$g_{\mu\nu}$ satisfying an equation to be specified shortly. 
A one dimensional string traces out a two dimensional worldsheet. So the
dynamics of the string is described by a two dimensional theory describing
how this worldsheet moves in spacetime. It is convenient to introduce
a metric on the worldsheet, $q_{ab}$. If $\s^a$ are local coordinates
on the worldsheet and $X^\mu$ are local coordinates on spacetime, the
dynamics of a string in this background is 
described by a two dimensional  sigma model:
\be 
S=\ell_s^{-2} \int d^2 \s \sqrt{-q} q^{ab} \nabla_a X^\mu \nabla_b X^\nu g_{\mu\nu}
\ee
Classically, this action is invariant under conformally rescaling the 
worldsheet metric, $q_{ab}$.
Quantum mechanically, there can be a conformal anomaly.
 We now demand that this conformal invariance
is preserved quantum mechanically. In other words, we have 
a two dimensional conformal field theory (CFT).
The vanishing of the conformal anomaly imposes an equation on the
spacetime metric, which becomes the field equation for $g_{\mu\nu}$.
 In a derivative expansion, this
looks like Einstein's equation at leading order, but then
has higher order terms involving powers and derivatives of the curvature
multiplied by appropriate powers of the string scale $\ell_s$. 

Associated with every two dimensional CFT is a number, $c$, called the
central charge. For a free field theory, $c$ is just the number of
scalar fields. So for a string in flat spacetime, $c$ is the spacetime
dimension. The presence of two dimensional diffeomorphism invariance
in string theory leads to the condition that $c=26$. So without fermions
and supersymmetry, string theory is consistent in 26 spacetime 
dimensions. If one adds fermions, one only needs 10 dimensions, as mentioned
earlier.
 We are thus lead to the following general definition of 
perturbative string theory:

 {\it Perturbative string theory
is equivalent to two dimensional conformal field theory with $c=26$.}

This is perturbative in the sense that different topologies for the
two dimensional worldsheet correspond to different orders in a quantum loop 
expansion. This is usually studied for static (or stationary) backgrounds
where one can analytically continue to Euclidean signature. Then one 
can work with Euclidean worldsheet metrics  and Riemann surfaces. Higher
genus surfaces describe higher loop contributions. The CFT
on $S^2$ describes the classical (tree level) theory, the CFT on $T^2$
describes the one loop correction, etc. Much less is known about this
quantum loop expansion when the backgrounds are time dependent.

Notice that the role of spacetime has  changed dramatically. The focus
is no longer on  fields propagating on spacetime, but rather 
on two dimensional quantum field theories. The spacetime metric acts like
``coupling constants" on the two dimensional fields.
This has several far reaching implications which we now discuss.

Since the string only senses the spacetime through this
sigma model, two metrics which
yield the same sigma model are indistinguishable in string theory. 
Apparently trivial changes to the sigma model action can result in
dramatic changes to the spacetime geometry. We give two examples.
The first is called T-duality \cite{Buscher:1987sk,Giveon:1994fu}.
Consider a metric with a Killing field
$\p/\p Y$. For simplicity, we will also assume there is a reflection symmetry
$Y\ra -Y$, so the metric is $ds^2 = g_{ij} dX^i dX^j + f dY^2$ where
$g_{ij}$ and $f$ can depend on $X^i$ but are independent of $Y$.
The two dimensional sigma model is
\be
S=\ell_s^{-2} \int   [g_{ij}\nabla_a X^i \nabla^a X^j + f\nabla_a Y\nabla^a Y]
\ee
The field equation for $Y$ is $\nabla_a (f\nabla^a Y)=0$ which implies
$d(f\ {}^*dY)=0$. Let us introduce a new field $\t Y$ via
$f{}^*dY = d\t Y$. Then the action becomes\footnote{Strictly speaking,
this ``derivation" of T-duality only works on euclidean worldsheets since
otherwise there is a minus sign in front of the second term in
(\ref{tdual}). A proper
derivation involves gauging the symmetry associated with the Killing field
and imposing a constraint that the associated field strength vanish 
\cite{Giveon:1994fu}.}
\be\label{tdual}
S=\ell_s^{-2} \int  [g_{ij}\nabla_a X^i \nabla^a X^j + f^{-1}
\nabla_a \t Y\nabla^a \t Y]
\ee
It turns out that the change of variables from $Y$ to $\t Y$ leaves the
CFT invariant \cite{Rocek:1991ps}. (More precisely, this is true if
$Y$ and $\t Y$ are both periodically identified with inverse periods.)
However (\ref{tdual})
describes a string moving in a geometrically different spacetime. 
We have replaced $g_{YY}$  with $ 1/g_{YY}$.
 In the simplest case of flat spacetime
with
one direction compactified on a circle, this shows that a circle
of radius $R$ is equivalent to  one with radius $\ell_s^2/R$.
Intuitively, the reason
is that strings in this background have two types of states. If the string
winds around the circle $n$ times, its energy is $nR/\ell_s^2$, while if it is
moving around the circle its energy is $m/R$ for some integer $m$
since momentum
is quantized. Clearly, if we interchange $R$ and $\ell_s^2/R$, and $m ,n$
the spectrum of states is unchanged. One can show that all string interactions
are also invariant.

Another example involves strings moving on spacetimes of the form $M_4\times
K$ where $K$ is a compact six dimensional space. The condition of
conformal invariance (and four dimensional supersymmetry) requires
that $K$ be Ricci flat and Kahler, i.e., a Calabi-Yau space
\cite{Candelas:1985en}.
By changing a sign (of a left moving
$U(1)$ charge) in the sigma model, one changes its interpretation from
a string moving on $K$ to a  string moving on a geometrically and topologically 
 different space, called the mirror of $K$. Since the sign is arbitrary
 from the CFT standpoint, 
strings on $K$ are equivalent to strings on $\tilde K$ 
\cite{Greene:1990ud}.  This 
``mirror symmetry"
has been checked in a dramatic way. By doing a calculation on $K$ and
reinterpreting it in terms of $\t K$, Candelas et al. \cite{Candelas:1990rm}
were able to calculate
the number of rational curves of degree $k$ in a particular Calabi-Yau
space $\t K$.
These are maps of
$S^2$ into $\t K$ described by equations of degree $k$. These numbers grow very
quickly and look like
\beq
k=1 \qquad & & 2875\cr
k=2 \qquad & &609,250\cr
k=3 \qquad & &317,206,375\cr
k=4 \qquad & &242,467,530,000
\eeq
These numbers
are very hard to calculate directly using traditional methods.  At the time
\cite{Candelas:1990rm} appeared, mathematicians had only been able to
calculate the first two. Since then,  new techniques were found to compute
them directly and they all agree with the predictions of mirror symmetry
\cite{GIV}.

Another consequence of the fact that strings see the spacetime only 
through the sigma model is that spacetimes which are singular in
general relativity can be completely nonsingular in string theory.
To see this, first note that the definition of a singularity in string
theory is different than in general relativity.
In general relativity, we usually  define a singularity in
terms of geodesic incompleteness which is based on the motion of test particles.
In string theory, we use the sigma model which describes test strings.
So a spacetime is considered 
singular if test strings are not well behaved\footnote{Strictly speaking,
one should also require that the  other objects in the 
 theory --branes -- have well behaved propagation.}.
A simple example  of a spacetime which is singular in general relativity
but not string theory is the quotient
of Euclidean space by a discrete subgroup of the rotation group. The
resulting space, called an orbifold, has a conical singularity at the
origin. Even though this leads to geodesic incompleteness in general 
relativity, it is completely harmless in string theory. This is essentially
because strings are extended objects. 

The orbifold has a very mild singularity,
but even curvature singularities can be harmless in string theory. 
A simple example follows from applying T-duality to rotations in the
plane. This results in the metric $ds^2 = dr^2 + (1/r^2) d\phi^2$
which has a curvature singularity at the origin. However strings on
this space are completely equivalent to strings in flat space.

As mentioned above, string theory has exact solutions which are the product
of four dimensional Minkowski space, and a compact Calabi-Yau
space. A given Calabi-Yau manifold usually admits a whole family of 
Ricci flat metrics. So one can construct a solution in which the four large
dimensions stay approximately flat and the geometry of the Calabi-Yau 
manifold changes slowly from one Ricci flat metric to another. In this process
the  Calabi-Yau space can develop a curvature singularity.
In many cases, this can be viewed as arising from 
a topologically nontrivial $S^2$ or $S^3$ being shrunk down
to zero area. 
It has been shown that when this happens, string theory remains completely well
defined. The evolution continues through the geometrical
singularity to a nonsingular Calabi-Yau space on the other side 
\cite{Aspinwall:1993nu,Strominger:1995cz}.

The reason this happens is roughly the following. There are extra degrees of
freedom in the theory
associated with branes wrapped around topologically 
nontrivial surfaces. As long as the area of the surface is nonzero, these
degrees of freedom are massive, and it is consistent to ignore them. However
when the surface shrinks to zero volume these degrees of freedom become
massless, and one must include them in the analysis. When this is done, 
the theory is nonsingular.

The above singularities are all a product of time and a singular space.
 However
other singularities which involve time in a crucial way have also been shown
to be harmless. Putting many branes on top of each other produces a 
gravitational field which often has a curvature singularity at the location
of the brane.  It has been shown that one can understand
physical processes near this
singularity 
in terms of excitations of the branes.

However it is simply not true that all singularities are removed in string
theory. Consider the plane wave:
\be\label{planewave}
ds^2 = -2dudv +dx_i dx^i + h_{ij}(u)x^i x^j du^2
\ee
If $h_{ij}$ is traceless, this metric is Ricci flat. The $u$ dependence
is arbitrary. It is easy to show that all the higher order perturbative
corrections to Einstein's equation vanish since the curvature is null
\cite{Horowitz:1989bv}. In fact, one can show that (\ref{planewave})
defines an exact conformal field theory \cite{Amati:1988sa}, so this
is an exact solution to string theory. 
If $h_{ij}$ diverges as $u\ra u_0$, then the plane wave is singular. One
can study string propagation in this background and show that
in some cases, the string does not have well behaved propagation
through this curvature singularity. The divergent tidal forces cause
the string to become infinitely excited \cite{Horowitz:1989bv}. 

Despite all this progress,
we still do not yet have a good understanding of the most important
types of  singularities: those arising from gravitational collapse
or cosmology. This remains an area of active investigation.

We have focused on the metric, but classical backgrounds for
string theory can also include
some matter fields. The massless states of the ten dimensional superstring
are in one-to-one correspondence with the fields of supergravity, so these
are the allowed matter fields. Their classical field equations look like
the supergravity equations plus higher derivative corrections.

We have mainly discussed ten dimensional backgrounds since this is
the critical dimension for the superstring. But there are ways
around this restriction. Recall that the fundamental requirement is
that the central charge $c$ is fixed.
 If the other
matter fields are allowed to be nonzero at infinity, the dimension of spacetime
can be reduced keeping the central charge unchanged. For example, if the dilaton grows linearly in a
spatial direction, then the dimension of spacetime can be reduced as low as two
\cite{Myers:1987fv}!
Also, if the three form $H=dB$ is nonzero at infinity, the dimension is also
changed. This arises naturally if the spacetime is a product of time and
a group manifold \cite{Gepner:1986wi}.

There are more subtle ways to lower the spacetime dimension.
One of these is called an asymmetric orbifold \cite{Narain:1986qm}.
Suppose one wants
a consistent background with $n<10$ dimensions. One starts with
a ten dimensional flat spacetime.
 Since the fields $X^\mu$ satisfy
a wave equation on the worldsheet,
they can be divided up into a left moving and right moving
part $X^\mu=X_L^\mu + X_R^\mu$. Now one takes $10-n$ of these fields
and makes different identifications on  $X_L$ and  $X_R$. It is as if
 $X_L$ and $X_R$ are living on different orbifolds. It clearly
no longer  makes sense to view these fields as coordinates on space. The
physical space is now lower dimensional, and the central charge is
made up by some fields on the worldsheet that do not have a spacetime
interpretation.

One can take this  idea even further.
So far, we have mainly discussed CFT's which come from a sigma model.
Even though spacetime has unusual properties, at least there is a spacetime.
These examples  can thus be viewed as the geometric phase of string theory.
However string theory is not restricted to sigma models. As we have seen,
one can consider any two dimensional conformal
field theory with $c=26$. Some of these can be viewed as describing
strings moving in spacetimes where the curvature is of order the string
scale and not really well defined. 

\setcounter{equation}{0}
\section{Nonperturbative lesson: Holography}

Our knowledge of nonperturbative string theory is still incomplete, but it
has already yielded a radical new view of spacetime called holography.
Roughly speaking, holography is the idea that physics in a region can be
described by fundamental degrees of freedom living on the boundary of
this region.  The idea that quantum gravity might be holographic
was first suggested by 't Hooft \cite{'tHooft:1993gx} and Susskind 
\cite{Susskind:1994vu} motivated by
the fact that black hole entropy is proportional to its horizon area.

The most concrete form of this idea is due to Maldacena \cite{Maldacena:1997re}
and called the
AdS/CFT correspondence (although it is really a conjecture). 
Maldacena considered a stack of $N$ parallel D 3-branes on top of each other.
As mentioned earlier, the strength of the gravitational field this produces
is governed by $g_s N$. When $g_s N\ll 1$ the spacetime is nearly flat and
there are two types of string excitations. There are open strings on the
brane whose low energy modes are described by a $U(N)$ gauge theory.
There are also closed strings away from the brane. When $g_s N\gg 1$, the
backreaction is important and the metric describes an extremal black 3-brane.
This is a generalization of a black hole appropriate for
a three dimensional extended object. It is extremal with respect to the
charge carried by the 3-branes, which sources the five form $F_5$.
Near the horizon, the spacetime becomes a product of $S^5$ and five
dimensional anti de Sitter (AdS) space.  (This is directly
analogous to the fact that near the horizon of an extremal Reissner-Nordstrom
black hole, the spacetime is $AdS_2\times S^2$.) String states near the
horizon are strongly redshifted and have very low energy  as seen 
asymptotically. In a certain  low energy
limit, one can decouple these strings from
the strings in the asymptotically flat region. At
weak coupling, $g_s N\ll 1$, this same limit decouples the closed strings
from the excitations of the 3-branes. Thus we get a connection between
a gauge theory at weak coupling, and strings in $AdS_5\times S^5$ at 
strong coupling. But both of these theories are (in principle) defined for all 
values of the coupling. Maldacena suggested that they were in fact
equivalent physical descriptions.
More precisely, 

{\it String theory with $AdS_5\times S^5$ boundary conditions
is equivalent to  a four dimensional ${\cal N}=4$ supersymmetric
$U(N)$ gauge theory.}

\noindent The gauge theory is conformally invariant and hence is a
four dimensional conformal field theory (CFT). More generally, 
starting with  other branes (or several different types of branes),
one is led to the following statement:  

{\it  AdS/CFT Correspondence: 
String theory on spacetimes which 
asymptotically approach the product of anti de Sitter (AdS) and a compact
space, is completely described by a conformal field theory 
``living on the
boundary at infinity".}

We will focus on the first form of this correspondence, since
this is the best understood case.
At first sight this conjecture seems unbelievable. 
How could an ordinary field theory
describe all of string theory? These two theories certainly look different
at weak coupling, but a crucial aspect of this correspondence is that when
the string theory is  weakly coupled, the gauge theory is strongly coupled
and vice versa. This is because the radius of curvature of $AdS_5$
and $S^5$ are both given by
\be\label{adsradius}
\ell = (4\pi g_s N)^{1/4} \ell_s
\ee
The Yang-Mills coupling $g_{YM}$ is related to the string theory
coupling by $g_{YM}^2 = 2\pi g_s$. The effective coupling in a large
$N$ gauge theory is the 't Hooft coupling, $g_{YM}^2 N$, so this must be
large in order for $\ell \gg\ell_s$ which is necessary for even a spacetime
interpretation to be valid in the string theory.

One sign that this is not completely crazy comes from
comparing the symmetries. 
 $\N=4$  $U(N)$ super Yang-Mills has a gauge field,
four (Weyl) fermions and six scalars $\vp^i$, all in the adjoint representation
of the gauge group. It has an $SO(4,2)$ symmetry
coming from conformal invariance, and an $SO(6)$ symmetry coming
from rotation of the scalars. This agrees with the geometric
symmetries of $AdS_5\times S^5$. Thus all spacetimes which asymptotically
approach $AdS_5\times S^5$ have an asymptotic symmetry group which 
agrees with the gauge theory.
If the gauge theory is on $S^3\times {\bf R}$
with the radius of the three-sphere also given by $\ell$, then the
global time translation in the bulk agrees with time translation in
the field theory and hence the energies of states in the field theory and
string theory should agree.

One cannot prove the AdS/CFT correspondence since we do not have an independent
nonperturbative definition of string theory to compare it to.
In fact, since the gauge theory is defined nonperturbatively,
one can view this as a nonperturbative and (mostly) background independent
definition of string theory. A background spacetime metric only enters
in the boundary conditions at infinity. Of course this only makes sense
if the gauge theory can reproduce what is known about string theory.
There are many checks one can make, and so far, the AdS/CFT correspondence
has survived all of them.
I will now describe some of these tests. (For a more complete discussion, see
\cite{Aharony:1999ti}.)

The initial checks concerned perturbations of $AdS_5\times S^5$.
It was shown that all linearized 
supergravity states (massless string modes)
have corresponding states in the gauge theory with
the same energy \cite{Witten:1998qj}.
It was also shown that some interactions agree 
\cite{Lee:1998bx}.
For a long time, it was difficult to give a precise description of the
massive excited string states in terms of the strongly coupled gauge theory.
However considerable
progress has been made recently for a class of states with large
angular momentum on the $S^5$ or $AdS_5$ \cite{Berenstein:2002jq,Gubser:2002tv}.
 In some cases, one
can even construct a two dimensional sigma model directly from the gauge
theory and show that it agrees (in a certain approximation) to the 
sigma model describing strings moving in $AdS_5\times S^5$ 
\cite{Tseytlin:2004cj}.

For perturbations of $AdS_5\times S^5$, one can reconstruct the background
spacetime from the gauge theory as follows.
Fields on $S^5$ can be decomposed into spherical
harmonics, which can be described as symmetric traceless tensors on ${\bf R}^6:$
$T_{i\cdots j} X^i \cdots X^j$. Restricted to the unit sphere one gets a basis
of functions.  Recall that the gauge theory has six scalars and the
SO(6) symmetry
of rotating the $\vp^i$. So the operators $T_{i\cdots j} \vp^i\cdots \vp^j$
give information about position on $S^5$. The field theory lives
on $S^3\times {\bf R}$ which can be viewed as the boundary at infinity of
$AdS_5$. So the only remaining direction is the radial direction.
This is believed to 
correspond to an energy scale in the gauge theory. Large radii correspond
to high energy or short distance in the gauge theory. For example,
a fundamental string stretched into a circle of large radius in $AdS_5$
corresponds to a thin flux tube in the gauge theory. The fact that the flux
tube naturally expands corresponds to the fundamental string collapsing
down to smaller radii.

Evidence for the AdS/CFT correspondence goes far beyond these perturbative
checks. Recently, a detailed map has been found between a
class of nontrivial asymptotically $AdS_5\times S^5$ supergravity solutions 
and a class of states in the gauge theory \cite{Lin:2004nb}.
These states and geometries both
preserve half of the supersymmetry of $AdS_5\times S^5$ itself. On the field
theory side, one restricts to fields that are independent of $S^3$ and hence
reduce to $N\times N$ matrices. In fact, all the states are created by just
one scalar field, so it can be described by  a single matrix model. This
theory can be quantized exactly and the states can be labeled by arbitrary
closed curves on a plane. On the gravity side, one
considers solutions to ten dimensional supergravity involving just the metric
and (selfdual) five form $F$. The field equations are  simply $dF=0$ and
\be\label{fieldeq}
R_{\mu\nu} = F_{\mu\alpha_1\cdots\alpha_4}{F_\nu}^{\alpha_1\cdots\alpha_4}
\ee
There exist a large class
of stationary solutions to (\ref{fieldeq}) which have an
$SO(4)\times SO(4)$ symmetry, and can be obtained by solving a linear equation.
These solutions are nonsingular, have no event horizons, but can have 
complicated topology. They are also labeled by arbitrary closed curves
on a plane. This provides a precise way to map states in the field
theory into bulk geometries \cite{Lin:2004nb}.
Only for some ``semi-classical" states
is the curvature below the Planck scale everywhere. It is natural to 
conclude that the full quantum gravity description of this sector of 
the theory is just given by the matrix model.

Understanding the Hawking-Bekenstein entropy of black holes
in terms of quantum states had been a longstanding problem. However
the AdS/CFT correspondence provides a natural solution. 
A  black hole in $AdS_5$ is described by the Schwarzschild AdS geometry
\be\label{adsbh}
ds^2 = -\({r^2\over \ell^2} +1 - {r_0^2\over r^2}\) dt^2 + \({r^2\over \ell^2}
+1 - {r_0^2\over r^2} \)^{-1}
dr^2 + r^2 d\Omega_3
\ee
Denoting the Schwarzschild radius by $r_+$, 
the Hawking temperature of this black hole is
\be
T_H = {\ell^2 + 2r_+^2\over 2\pi r_+\ell^2}
\ee
When $r_+ \gg\ell$, the Hawking
temperature is large, $T_H \sim r_+/\ell^2$. 
This is quite different from a large
black hole in asymptotically flat spacetime which has $T_H \sim 1/r_+$.
The gauge theory
description is just a thermal state at the same temperature $T_H$.
Let us compare the entropies. It is difficult to calculate the field
theory entropy at strong coupling, but at weak coupling, we have of
order $N^2$ degrees of freedom, on a three sphere of radius $\ell$ at 
temperature $T_H$ and hence 
\be\label{entropy}
S_{YM} \sim N^2 T_H^3 \ell^3.
\ee
On the string theory side, the solution is the product of (\ref{adsbh})
and an $S^5$ of radius $\ell$. 
So recalling that $G\sim g_s^2\ell_s^8$ in ten dimensions and dropping factors
of order unity, the Hawking-Bekenstein entropy of this black hole is
\be 
S_{BH}={A\over 4G}\sim {r_+^3 \ell^5\over g_s^2 \ell_s^8}\sim {T_H^3\ell^{11}
\over g^2_s\ell_s^8}
\sim N^2 T_H^3 \ell^3
\ee
where we have used (\ref{adsradius}) in the last step. The agreement
with (\ref{entropy}) shows that the field theory has enough states to
reproduce the entropy of large black holes in $AdS_5$. Putting in all
the numerical factors one finds that $S_{BH} = {3\over 4} S_{YM}$ 
\cite{Gubser:1996de}. Since
$S_{BH}$  is a measure of the number of states at strong coupling, and
$S_{YM}$ has been calculated at weak coupling, it should not be surprising
that they are not precisely equal.  We do not yet understand
why they are related by a simple factor of $3/4$.

There is another test one can perform with
 the gauge theory at finite temperature. At long wavelengths, one can
use a hydrodynamic approximation and think of this as a fluid. It is
then natural to ask:
What is the speed of sound waves? Conformal invariance implies that
the stress energy tensor is traceless, so $p=\rho/3$ which implies
that $v = 1/\sqrt 3$. The question is: Can you derive this sound speed
from the AdS side? This would seem to be difficult since the bulk does not
seem to have any preferred speed other than the speed of light. But recent
work has shown that the answer is yes \cite{Policastro:2002tn}. 
Using the correspondence, one can also compute other hydrodynamic quantities
such as the shear viscosity, but these are hard to check since they
are difficult to calculate directly
in the strongly coupled thermal gauge theory. 
There is also a field theory interpretation
of black hole quasinormal modes. A perturbation of the black hole decays with a
characteristic time set by the imaginary part of the lowest quasinormal 
mode. This should correspond to the timescale for the gauge theory to return
to thermal equilibrium. One can show that the quasinormal mode frequencies
are poles in the retarded Green's function of a certain operator in the gauge
theory.
The particular operator depends on the type of field used to perturb the
black hole.

In quantum field theory
there is a standard procedure for integrating out high energy
degrees of freedom and obtaining an effective theory at low energy.
This is known as renormalization group (RG) flow. If one starts with a 
conformal field theory at high energy, the RG flow is trivial. The low
energy theory looks the same as the high energy theory. This is because
there is no intrinsic scale. But if you perturb the theory by adding mass terms
to certain fields, the RG flow is nontrivial and one obtains a different
theory at low energies. Since the energy scale corresponds to radius, this
RG flow in the boundary field theory 
corresponds to radial dependence in the bulk.
Turning on mass terms corresponds to changing the boundary conditions for 
certain matter fields in the bulk. These new boundary conditions require
that the matter fields are nonzero, so AdS is no longer a solution.
 By solving Einstein equation with these new boundary conditions,
one obtains a solution which approaches  AdS (with, in general, a different
radius of curvature) at small radius. By comparing
the small $r$ behavior with the endpoint of the RG flow one finds detailed
numerical agreement \cite{Freedman:1999gp}. So the classical Einstein
equation knows a lot about RG flows in quantum field theory! Since we start by
adding mass terms to the field theory, this also shows that the AdS/CFT
correspondence can be extended to some nonconformal field theories. It is
not yet clear how many quantum field theories have dual descriptions
in terms of string theory. It has been suggested that this should be true
for all theories with a large $N$ limit.

\setcounter{equation}{0}
\section{Discussion}

It should be clear that conformal field theories are playing a central role
in our current understanding of string theory. Two dimensional CFT's with the
right central charge describe classical string solutions and the quantum
loop expansion. Other CFT's with (typically)
different spacetime dimension are believed to
provide a complete nonperturbative description of string theory with
asymptotically AdS boundary conditions. There is even one version
of the AdS/CFT correspondence in which the dual CFT is two dimensional.
This arises when the spacetime is asymptotically $AdS_3\times S^3\times T^4$.
One thus has the confusing situation that a 2D CFT describes
perturbative string excitations about this space, and a different 2D CFT
describes the entire theory. What is even more surprising is that
the first CFT describes loop corrections by changing the space it
is defined on to higher genus Riemann surfaces, while the second CFT
provides a complete nonperturbative description keeping the space it
is defined on fixed (just $S^1\times {\bf R}$)! In this case one again has
a detailed map between certain supersymmetric states in the (second) CFT
and nontrivial gravity solutions \cite{Lunin:2002bj}.
This case has at least one advantage
over the $AdS_5\times S^5$ case discussed above. The entropy of large
black holes
can now be reproduced exactly, including the numerical coefficient. This is 
related to the fact that a black hole in $ AdS_3$ is a BTZ black hole
which is locally $ AdS_3$ everywhere. Thus when one extrapolates to small
coupling, one does not modify the geometry with higher curvature corrections.

There are other approaches to nonperturbative string theory such as
string field theory. 
I have focused on the AdS/CFT correspondence since that is both the most
extensively studied, and
also contains the most far reaching lesson about the nature of spacetime. 
It provides answers to some longstanding questions about quantum gravity.
For example, it has often been suggested that 
 space and time should
be derived quantities in quantum gravity. But the problem has always been: 
If space and time are not fundamental, what replaces them? Here the answer
is that there is an auxiliary spacetime metric which is fixed by the 
boundary conditions at infinity. The CFT uses this metric, but the physical
spacetime metric is a derived quantity. It is important to emphasize that
the spacetime is not emerging from ``strings". In this approach, the
so-called fundamental strings of string theory are also derived quantities.
Both the strings and spacetime are constructed from the CFT.

As another example,  consider the formation
and evaporation of a small black hole in a spacetime which is asymptotically
$AdS_5\times S^5$. By the AdS/CFT correspondence, this process is described
by ordinary unitary evolution in the CFT. So black hole evaporation 
does not violate quantum mechanics. 

However there remain many open questions. 
The dictionary relating spacetime
concepts in the bulk and field theory concepts on the boundary
is very incomplete, and still being
developed.  For example, while we know how to translate certain states of
the CFT into bulk geometries, we do not yet know the general condition on the
state in order for a semiclassical spacetime to be well defined.
Another class of questions concerns how to formulate holography for
other boundary conditions. 
Can holography be extended to asymptotically flat or cosmological
spacetimes? If so, will the dual description again be given in terms of a 
local quantum field theory?
One extension that has already been carried out is to plane wave spacetimes.
Penrose has shown that every spacetime has a plane wave as a limit. 
The idea is to blow up the geometry in a small neighborhood of a
null geodesic. Consider a null geodesic which stays at the origin
of the $AdS_5$ but circles around the equator of the $S^5$. Taking
the Penrose limit yields a particularly simple plane wave
\be
ds^2 = -2dudv +dx_i dx^i - x^2 du^2
\ee
which has an $SO(8)$ symmetry. Berenstein et al. \cite{Berenstein:2002jq}
showed what this Penrose limit
corresponds to in the dual gauge theory. In this way, one can
extend the AdS/CFT correspondence to strings in a plane wave background.
One advantage is that excited string states can now be studied more
easily and the entire string spectrum has been shown to agree with  states
in the
dual theory.
Even some interactions have been shown to agree \cite{Pearson:2002zs}.

The AdS/CFT correspondence can also be used to gain information about
strongly coupled gauge theories. Certain calculations are  easier
to do on the gravity  side and then translated into new field theory
results. Although this has not been our
focus here, there has been considerable effort in this direction
motivated by a desire to better understand the strong interactions.
Already, people have found geometrical analogs of confinement and
chiral symmetry breaking \cite{Klebanov:2000hb}.

The picture of spacetime emerging from string theory certainly
seems bizarre and unconventional. But the notion of curved spacetime
must have seemed equally bizarre and unconventional to the physicists of
the early twentieth century.
It is clear that we do not yet have the final story. 
The picture is still emerging. Hopefully we will have
the answer well before the $200^{th}$ anniversary of spacetime.
One can hardly imagine what physics will look like then.

\vskip 1cm

\centerline{\bf Acknowledgments}
\vskip .5cm
I would like to thank O. Aharony and D. Marolf for discussions which helped
clarify the above presentation.
This work was supported in part by NSF grant PHY-0244764.


\begin{thebibliography}{99}

\bibitem{Ashtekar}
A. Ashtekar, contribution to this volume.

\bibitem{Marolf:2003tx}
D.~Marolf,
``Resource letter: The nature and status of string theory,''
Am.\ J.\ Phys.\  {\bf 72} (2004) 730
[arXiv:hep-th/0311044].


\bibitem{POL} J. Polchinski, {\it String Theory}, in 2 vols., Cambridge Univ.
Press, 1998.

\bibitem{ZEI} B. Zweibach, {\it A First Course in String Theory}, Cambridge Univ. Press, 2004.

\bibitem{Polchinski:1995mt}
J.~Polchinski,
``Dirichlet-Branes and Ramond-Ramond Charges,''
Phys.\ Rev.\ Lett.\  {\bf 75} (1995) 4724
[arXiv:hep-th/9510017].

\bibitem{Arkani-Hamed:1998rs}
N.~Arkani-Hamed, S.~Dimopoulos and G.~R.~Dvali,
``The hierarchy problem and new dimensions at a millimeter,''
Phys.\ Lett.\ B {\bf 429} (1998) 263
[arXiv:hep-ph/9803315].

\bibitem{Randall:1999vf}
L.~Randall and R.~Sundrum,
``An alternative to compactification,''
Phys.\ Rev.\ Lett.\  {\bf 83} (1999) 4690
[arXiv:hep-th/9906064].

\bibitem{Emparan:2001wn}
R.~Emparan and H.~S.~Reall,
``A rotating black ring in five dimensions,''
Phys.\ Rev.\ Lett.\  {\bf 88} (2002) 101101
[arXiv:hep-th/0110260].

\bibitem{Elvang:2004rt}
H.~Elvang, R.~Emparan, D.~Mateos and H.~S.~Reall,
``A supersymmetric black ring,''
arXiv:hep-th/0407065.

\bibitem{Wiseman:2002zc}
T.~Wiseman,
``Static axisymmetric vacuum solutions and non-uniform black strings,''
Class.\ Quant.\ Grav.\  {\bf 20} (2003) 1137
[arXiv:hep-th/0209051].


\bibitem{Buscher:1987sk}
T.~H.~Buscher,
``A Symmetry Of The String Background Field Equations,''
Phys.\ Lett.\ B {\bf 194} (1987) 59.


\bibitem{Giveon:1994fu}
A.~Giveon, M.~Porrati and E.~Rabinovici,
``Target space duality in string theory,''
Phys.\ Rept.\  {\bf 244} (1994) 77
[arXiv:hep-th/9401139].

\bibitem{Rocek:1991ps}
M.~Rocek and E.~Verlinde,
``Duality, quotients, and currents,''
Nucl.\ Phys.\ B {\bf 373} (1992) 630
[arXiv:hep-th/9110053].

\bibitem{Candelas:1985en}
P.~Candelas, G.~T.~Horowitz, A.~Strominger and E.~Witten,
``Vacuum Configurations For Superstrings,''
Nucl.\ Phys.\ B {\bf 258} (1985) 46.

\bibitem{Greene:1990ud}
B.~R.~Greene and M.~R.~Plesser,
``Duality In Calabi-Yau Moduli Space,''
Nucl.\ Phys.\ B {\bf 338} (1990) 15;
P.~Candelas, M.~Lynker and R.~Schimmrigk,
``Calabi-Yau Manifolds In Weighted P(4),''
Nucl.\ Phys.\ B {\bf 341} (1990) 383.

\bibitem{Candelas:1990rm}
P.~Candelas, X.~C.~De La Ossa, P.~S.~Green and L.~Parkes,
``A Pair Of Calabi-Yau Manifolds As An Exactly Soluble Superconformal
Theory,''
Nucl.\ Phys.\ B {\bf 359} (1991) 21.

\bibitem{GIV}
A. Givental, ``The Mirror Formula for Quintic Threefolds", math.AG/9807070.

\bibitem{Aspinwall:1993nu}
P.~S.~Aspinwall, B.~R.~Greene and D.~R.~Morrison,
``Calabi-Yau moduli space, mirror manifolds and spacetime topology  change in
string theory,''
Nucl.\ Phys.\ B {\bf 416} (1994) 414
[arXiv:hep-th/9309097].


\bibitem{Strominger:1995cz}
A.~Strominger,
``Massless black holes and conifolds in string theory,''
Nucl.\ Phys.\ B {\bf 451} (1995) 96
[arXiv:hep-th/9504090];
B.~R.~Greene, D.~R.~Morrison and A.~Strominger,
``Black hole condensation and the unification of string vacua,''
Nucl.\ Phys.\ B {\bf 451} (1995) 109
[arXiv:hep-th/9504145].

\bibitem{Horowitz:1989bv}
G.~T.~Horowitz and A.~R.~Steif,
``Space-Time Singularities In String Theory,''
Phys.\ Rev.\ Lett.\  {\bf 64} (1990) 260.

\bibitem{Amati:1988sa}
D.~Amati and C.~Klimcik,
``Nonperturbative Computation Of The Weyl Anomaly For A Class Of Nontrivial
Backgrounds,''
Phys.\ Lett.\ B {\bf 219} (1989) 443.


\bibitem{Myers:1987fv}
R.~C.~Myers,
``New Dimensions For Old Strings,''
Phys.\ Lett.\ B {\bf 199} (1987) 371.

\bibitem{Gepner:1986wi}
D.~Gepner and E.~Witten,
``String Theory On Group Manifolds,''
Nucl.\ Phys.\ B {\bf 278} (1986) 493.

\bibitem{Narain:1986qm}
K.~S.~Narain, M.~H.~Sarmadi and C.~Vafa,
``Asymmetric Orbifolds,''
Nucl.\ Phys.\ B {\bf 288} (1987) 551.

\bibitem{'tHooft:1993gx}
G.~'t Hooft,
``Dimensional reduction in quantum gravity,''
arXiv:gr-qc/9310026.


\bibitem{Susskind:1994vu}
L.~Susskind,
``The World as a hologram,''
J.\ Math.\ Phys.\  {\bf 36} (1995) 6377
[arXiv:hep-th/9409089].


\bibitem{Maldacena:1997re}
J.~M.~Maldacena,
``The large N limit of superconformal field theories and supergravity,''
Adv.\ Theor.\ Math.\ Phys.\  {\bf 2} (1998) 231
[Int.\ J.\ Theor.\ Phys.\  {\bf 38} (1999) 1113]
[arXiv:hep-th/9711200].


\bibitem{Aharony:1999ti}
O.~Aharony, S.~S.~Gubser, J.~M.~Maldacena, H.~Ooguri and Y.~Oz,
``Large N field theories, string theory and gravity,''
Phys.\ Rept.\  {\bf 323} (2000) 183
[arXiv:hep-th/9905111].

\bibitem{Witten:1998qj}
E.~Witten,
``Anti-de Sitter space and holography,''
Adv.\ Theor.\ Math.\ Phys.\  {\bf 2} (1998) 253
[arXiv:hep-th/9802150].

\bibitem{Lee:1998bx}
S.~M.~Lee, S.~Minwalla, M.~Rangamani and N.~Seiberg,
``Three-point functions of chiral operators in D = 4, N = 4 SYM at  large N,''
Adv.\ Theor.\ Math.\ Phys.\  {\bf 2} (1998) 697
[arXiv:hep-th/9806074].


\bibitem{Berenstein:2002jq}
D.~Berenstein, J.~M.~Maldacena and H.~Nastase,
``Strings in flat space and pp waves from N = 4 super Yang Mills,''
JHEP {\bf 0204} (2002) 013
[arXiv:hep-th/0202021].

\bibitem{Gubser:2002tv}
S.~S.~Gubser, I.~R.~Klebanov and A.~M.~Polyakov,
``A semi-classical limit of the gauge/string correspondence,''
Nucl.\ Phys.\ B {\bf 636} (2002) 99
[arXiv:hep-th/0204051].

\bibitem{Tseytlin:2004cj}
A.~A.~Tseytlin,
``Semiclassical strings and AdS/CFT",
arXiv:hep-th/0409296.

\bibitem{Lin:2004nb}
H.~Lin, O.~Lunin and J.~Maldacena,
``Bubbling AdS space and 1/2 BPS geometries,''
arXiv:hep-th/0409174.

\bibitem{Gubser:1996de}
S.~S.~Gubser, I.~R.~Klebanov and A.~W.~Peet,
``Entropy and Temperature of Black 3-Branes,''
Phys.\ Rev.\ D {\bf 54} (1996) 3915
[arXiv:hep-th/9602135].

\bibitem{Policastro:2002tn}
G.~Policastro, D.~T.~Son and A.~O.~Starinets,
``From AdS/CFT correspondence to hydrodynamics. II: Sound waves,''
JHEP {\bf 0212} (2002) 054
[arXiv:hep-th/0210220].

\bibitem{Freedman:1999gp}
D.~Z.~Freedman, S.~S.~Gubser, K.~Pilch and N.~P.~Warner,
``Renormalization group flows from holography supersymmetry and a  c-theorem,''
Adv.\ Theor.\ Math.\ Phys.\  {\bf 3} (1999) 363
[arXiv:hep-th/9904017].

\bibitem{Lunin:2002bj}
O.~Lunin, S.~D.~Mathur and A.~Saxena,
``What is the gravity dual of a chiral primary?,''
Nucl.\ Phys.\ B {\bf 655}, 185 (2003)
[arXiv:hep-th/0211292];
O.~Lunin, J.~Maldacena and L.~Maoz,
``Gravity solutions for the D1-D5 system with angular momentum,''
arXiv:hep-th/0212210.

\bibitem{Pearson:2002zs}
J.~Pearson, M.~Spradlin, D.~Vaman, H.~Verlinde and A.~Volovich,
``Tracing the string: BMN correspondence at finite J**2/N,''
JHEP {\bf 0305} (2003) 022
[arXiv:hep-th/0210102].


\bibitem{Klebanov:2000hb}
I.~R.~Klebanov and M.~J.~Strassler,
``Supergravity and a confining gauge theory: Duality cascades and
chiral symmetry breaking resolution of naked singularities,''
JHEP {\bf 0008} (2000) 052
[arXiv:hep-th/0007191].


\end{thebibliography}
\end{document}